\documentclass[11pt]{article}
\usepackage{hyperref}
\pdfoutput=1
\begin{document}
\title{Liquid Droplet Impact Dynamics on Micro-Patterned Superhydrophobic Surfaces}
\author{Cristian Clavijo, Daniel Maynes, Julie Crockett \\
\\\vspace{6pt} Department of Mechanical 
Engineering, \\ Brigham Young University, Provo, UT 84606, USA}
\maketitle
\section{Fluid Dynamics Video Description}

This fluid dynamics video (\href{http://flyht.byu.edu}{Liquid Droplet Impact Dynamics on Micro-Patterned Superhydrophobic Surfaces}) exhibits experimental qualitative and quantitative results of water/glycerol ($50\%/50\%$ by mass) droplet impact on two types of micro-patterned superhydrophobic surfaces.  The two types of surfaces used were $80\%$ cavity fraction (structured such that the liquid is in contact with $20\%$ of the surface) ribs and posts with a periodic spacing of $40$ microns and $32$ microns, respectively.  All surfaces were manufactured through photolithography.  The impact Weber number is defined as $We= \rho V^2 D/ \sigma$, where $\rho$ is the average density of the water/glycerol mixture, $V$ is the impact velocity, $D$ is the droplet diameter before impact, and $\sigma$ is the surface tension.  Weber number is used as the dynamic parameter to compare droplet splashing after impact and subsequent rebound behaviors between the two types of surfaces.  While droplets exhibit similar dynamics on both rib and post structured surfaces at low Weber numbers, rebound jet speed (normalized by droplet impact speed) is notably higher on posts than ribs for all Weber numbers tested here ($5<We<900$).  At elevated Weber numbers, droplets splash peripherally.  These side droplets created at splash are referred to as satellite droplets.  Top view videos show the significant difference in splashing behavior for the two surface types.  On ribs, satellite droplet formation consistently prefers a path $60^\circ$ from the ribs longitudinal direction.  This behavior prevails for Weber numbers greater than $150$, however, distinguished splashing does not occur for $We > 265$.  On posts, satellite droplets also follow a specific path but a different orientation from splashing on ribs.  Satellite droplets form in locations aligned with the post lattice structure.  This behavior is observed for $600 < We < 750$.

The rebounding jet, which occurs as the droplet comes back together after impact and splashing, exhibits an interesting phenomenon on ribs under certain conditions.  Due to the uneven shear distribution on the retracting droplet due to the surface anisotropy, a jet exhibiting two-prongs is observed.  This behavior occurs for $115 < We < 265$ on ribs \cite{PMW12}.  Two-pronged jets are also observed on posts but are less coherent and less repetitive.  This behavior was observed for posts around $500 < We < 600$.


\end{document}